\documentclass{ws-ijmpd}
\usepackage[super,compress]{cite}
\usepackage{pstricks}
\begin{document}


%
\catchline{}{}{}{}{}
%

\title{Gaussian black holes in Rastall Gravity }
\author{EURO SPALLUCCI}

\address{Department of Physics, University of Trieste, Strada Costiera 11\\
Trieste, Italy 34151,
Italy, and\\
I.N.F.N. Sezione di Trieste, Strada Costiera 11\\
Trieste, Italy 34151,
Italy
\\
euro@ts.infn.it}

\author{ANAIS SMAILAGIC}

\address{ I.N.F.N. Sezione di Trieste, Strada Costiera 11\\
Trieste, Italy 34151,
Italy\\
anais@ts.infn.it}

\maketitle

\begin{history}
\received{Day Month Year}
\revised{Day Month Year}
\end{history}

\begin{abstract}

In this short note we present the solution of Rastall gravity equations sourced by a Gaussian matter distribution. 
We find that the black hole metric shares all the common features 
of other regular, General Relativity BH solutions discussed in the literature: 
 there is no curvature singularity and the Hawking radiation leaves a remnant at zero temperature
in the form of a massive ordinary particle.

\end{abstract}


\ccode{  PACS numbers: 04.70.Dy,04.70.Bw,04.60.Bc}

\tableofcontents

\section{Introduction}	
In a variety of papers
\cite{Bar68,AyG00,Dym92,Dym02,Dym03,BrF06,BMD07,BrD07,Hay06,Ans08,FMM89,FMM90,BaP90}, 
regular solutions of Einstein field equations
generated by non-point like distributions of matter have been presented. Apart from different choices of matter 
source they all share the same characteristic features:
\begin{itemize}
	\item the curvature singularity at the origin is replaced by an inner core of regular de Sitter vacuum.\\
	\item They posses inner and outer horizons and admit an extremal configuration.\\
	\item The end-point of the Hawking evaporation is a zero temperature massive remnant.
\end{itemize}
Recently, a black hole solutions sourced by  a Gaussian energy momentum tensor have been 
considered in the framework of Rastall gravity \cite{Ma:2017jko}. Despite the regularity of the source, the authors argued  
that, in this particular case, the above features appears not to be simply realized. More in detail, they find that:
\begin{enumerate}
 \item there is no de Sitter core;
\item  the curvature singularity is still present in $r=0$;
\item  the tangential pressure diverges near $r=0$;
\item  the BH will totally convert its mass into thermal radiation leaving no remnant. 
\end{enumerate}
In this note we shall critically  analyze the above statements and
show that Gaussian BHs  shares all the characteristics of regular BHs contrary to the conclusions in \cite{Ma:2017jko}.\\
The paper is organized as follows. In Sect.(\ref{Rastg}) we present the essential features of Rastall gravity; in Sect.(\ref{RBH})
we solve the field equations with a Gaussian source. We discuss the regularity of the solution and its thermodynamical behavior.
In Sect.(\ref{end}) we summarize and the main results and their robustness.

\section{A short introduction to Rastall gravity}
\label{Rastg}
Rastall introduced  an interesting modification of General Relativity  \cite{Rastall:1973nw}, 
where the covariant conservation condition $T^{\mu\nu}_{\ ;\,\nu}=0$ is relaxed to a more general relation relation

\begin{equation}
 T^{\mu\nu}_{\ ;\,\nu} = a^\mu  \label{r1}
\end{equation}

Consistency with Special Relativity requires $a^\mu\to 0$ in the limit of vanishing space-time curvature.
Thus, a convenient choice, compatible with this limit, for a four-vector $ a^\mu$  is 

\begin{equation}
 a^\mu = \lambda R^{\ ,\,\mu}  \label{r2}
\end{equation}

where, $R$ is the Ricci scalar and $\lambda$ is a free parameter \footnote{In order to avoid misunderstanding, please, note
that $\lambda$ \emph{is not} the cosmological constant, but a new free parameter.}. \\
Equations (\ref{r1}), (\ref{r2}) lead to a  modified  Einstein field equations which read

\begin{eqnarray}
&& R_{\mu\nu}-\frac{1}{2}\left(\, 1 -2\kappa\lambda\,\right) g_{\mu\nu}R=\kappa T_{\mu\nu}\ ,\label{E1}\\
&& R=\frac{\kappa}{\left(4\,\kappa\,\lambda-1\right)}\,T\ ,\quad \kappa\Lambda\ne 1/4 \label{E2}\\
&& T^{\mu\nu}_{\ ,\,\nu}=\lambda R^{\ ;\,\mu} \label{E3}
\end{eqnarray}
It is clear that Eq.{\ref{E2}} is not defined for $\kappa\lambda =1/4 $. In this case, Eq.(\ref{E1}) reads

\begin{equation}
 R_{\mu\nu}-\frac{1}{4}g_{\mu\nu}R=\kappa T_{\mu\nu}\longrightarrow T^\mu_\mu =0 \ ,\qquad\forall R
\end{equation}

Thus, the choice $\kappa\lambda =1/4$ leads to a consistent set of field equations only for a traceless energy
momentum tensor. Even if this can be an interesting problem by itself, it is not the one we are going to address
in this note.\\
In order to reproduce Einstein equations in the limit  $\lambda\rightarrow 0$, the gravitational coupling 
$\kappa$ must be identified
as $\kappa =8\pi G_N$, but to make the field equations (\ref{E1}), (\ref{E2}), (\ref{E3}) analytically solvable, 
the better choice is  $\kappa\lambda=1/2$, $\kappa=4\pi G_N$. 
\begin{eqnarray}
&& R_{\mu\nu}=\kappa T_{\mu\nu}\ ,\label{E11}\\
&& T^{\mu\nu}_{\ ;\,\nu}=\frac{1}{2} T^{\ ,\,\mu} \label{E33}
\end{eqnarray}

where $T\equiv T^\mu_\mu$.\\
A Lagrangian formulation  for Rastall gravity with general $\lambda$ has been proposed
in \cite{Santos:2017nxm} and we refer to this paper for more details.\\
This generalized gravitational model has been recently used as an alternative, phenomenologically motivated, description 
in different cases \cite{deMello:2014hra,Oliveira:2015lka,Bronnikov:2017pmz,Moradpour:2016ubd,Heydarzade:2016zof}.\\

\section{Rastall Gaussian BHs}
\label{RBH}

In this Section we are going to look for the exact regular solutions of the field equations for Rastall gravity, 
in order to compare it to the known solutions in  General Relativity.  
The  source is given by an anisotropic fluid energy-momentum tensor
$T^\mu_\nu =Diag\left(\, -\rho\ , p_r\ , p_\perp\ , p_\perp\,\right) $, with matter density given by a Gaussian source
\begin{equation}
\rho\equiv \frac{M}{\left(\, 4\,\pi l_0^2\,\right)^{3/2}}e^{-r^2/4l_0^2}\label{sorg}
\end{equation}
$M$ is the total mass/energy of the system, given by 
\begin{equation}
 M\equiv 4\pi \int_0^\infty dr r^2 \rho(r)\ ,
\end{equation}
and the width $l_0$ of the Gaussian distribution is kept as a free parameter. In the original paper, where Gaussian BHs
were first introduced \cite{Nicolini:2005vd}, length scale $l_0$ was related to the parameter $\theta$ characterizing  
coordinate non-commutativity, but other choices for $l_0$ are equally possible 
\footnote{ The most ``traditional'' choice is probably $l_0=l_{Pl}$, but other assignments are equally physically
meaningful \cite{Spallucci:2017aod}. }\\
The fluid satisfies  the equation of state 
\begin{equation}
 p_r=-\rho\ ,
\end{equation}
which at short distance reproduces the ``\emph{vacuum}'' state equation $\rho(0)= -p_r(0)< \infty $\footnote{
It is important to remark that this is \emph{not a classical vacuum}. Rather, this state equation emulates the 
properties  of some possible ``quantum gravitational'' vacuum at short distance.
 The negative pressure provides a stabilizing mechanism avoiding the complete collapse of the matter
source and replaces the central singularity with a de Sitter core. }.\\
The generalized conservation (\ref{E2}) determines the remaining tangential pressure $p_\perp$  as
\begin{equation}
 \frac{dp_r}{dr}+\frac{2}{r} \left(\, p_r - p_\perp\,\right) 
=\frac{1}{2}\frac{d}{dr}\left(\, -\rho + p_r + 2p_\perp\,\right) \label{ptang}
\end{equation}
This equation will be solved in the next section.\\
We are looking for a spherically symmetric metric of the general form $ds^2 =-f(r)\,dt^2 +f(r)^{-1}\,dr^2 
+ r^2d\Omega^2 $, $f(r)\to 1$ as $r\to \infty$. 
The function $f$ is determined by the above matter source. One has to solve the single field equation

\begin{equation}
R_r^r={\kappa} T^r_r=   -\kappa\,\rho
\end{equation}

Oddly enough, since the $R^r_r$ component of the Ricci tensor is

\begin{equation}
 R_r^r=-\frac{1}{2\,r^2}\frac{d}{dr}\left(\,r^2\,f^\prime\, \right)\ ,
\end{equation}
it follows that the Rastall field equations  collapse to the classical Poisson equation

\begin{equation}
 \Delta f(r)= 2\kappa \rho
\end{equation}
\begin{figure}[h!]
\begin{center}
\includegraphics[width=11cm]{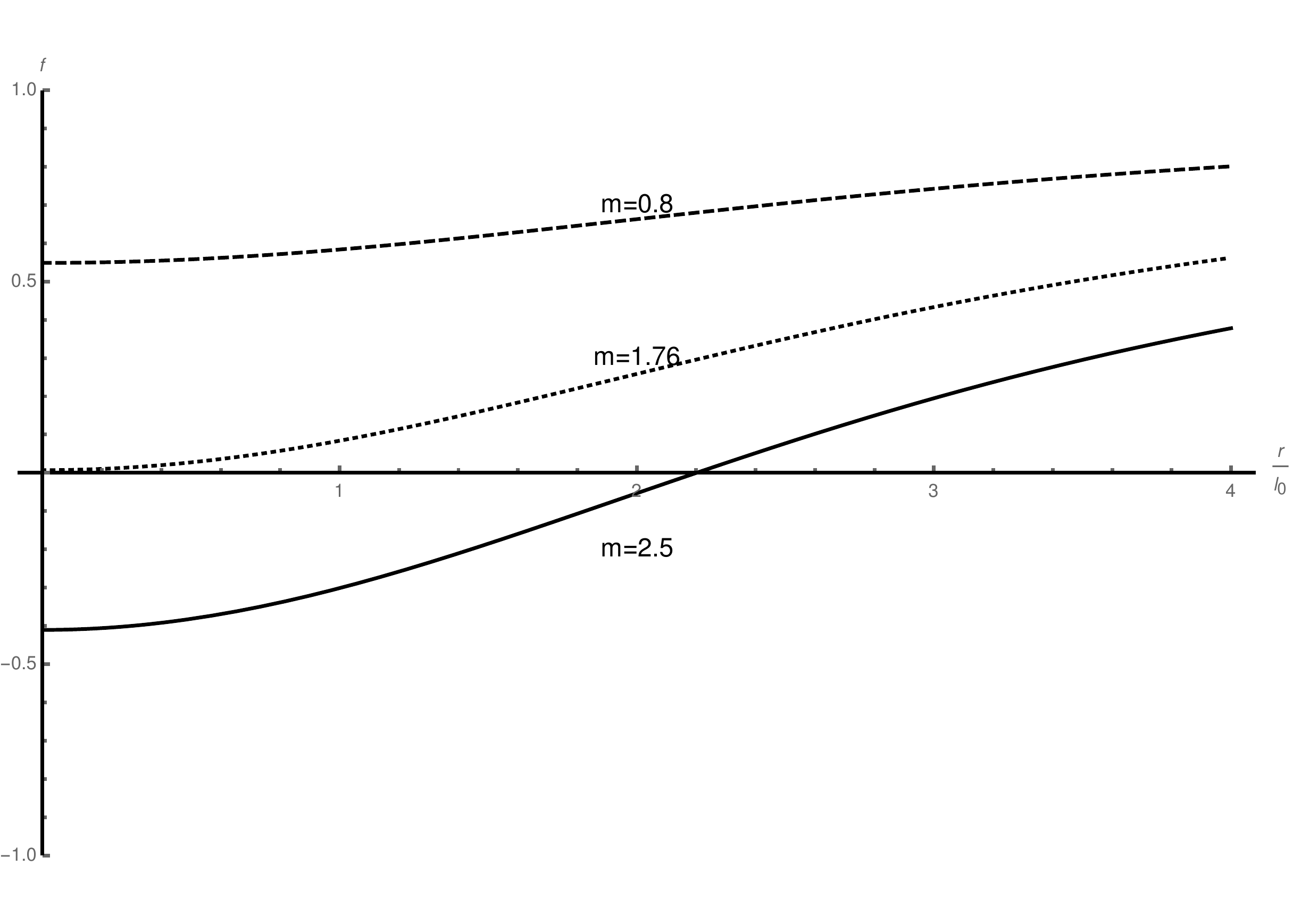}
\caption{Plot of the solution (\ref{rastbh}) for different masses  $m=M_{Pl} l_0/l_{Pl}$. Only 
one horizon exists for $M\ge M_0$,  the limiting mass $M_0=1.76 M_{Pl} l_0/l_{Pl} $. }
\end{center}
\end{figure}
which has the  known solution \cite{Nicolini:2005zi}

\begin{equation}
f(r)= 1 -\frac{2MG_N}{r}\frac{\gamma\left(\, 1/2\ ; r^2/4l_0^2\,\right)}{\Gamma(1/2)}
 \label{rastbh}
\end{equation}

where, $\gamma\left(\, 1/2\ ; r^2/4l_0^2\,\right)$ is the lower incomplete gamma-function  It is important to recall
that Gaussian BH solutions in 4D General Relativity have, instead, $\gamma\left(\, 3/2\ ; r^2/4l_0^2\,\right)$ 
\cite{Nicolini:2005vd,Ansoldi:2006vg, Nicolini:2008aj}. The presence of  $\gamma\left(\, 1/2\ ; r^2/4l_0^2\,\right)$ 
is a particular feature of the Rastall model and \emph{has nothing to do with any weak-field approximation, 
or classical limit} of the field equations.\\

In Sect.(\ref{RBH}) we study the short-distance behavior of the solution (\ref{rastbh}) and prove the existence of the
de Sitter core. We solve equation (\ref{ptang}) and show that $p_\perp$ is \emph{regular} everywhere. Finally, we study
the Hawking process and find that the evaporation ends at zero temperature leaving a cold, massive, particle as a finite
remnant of the original BH. Deviations from the standard ``area law'', due to the extended nature of the source, are also
computed in the ``large'' and ``small'' BH limits.\\
In Sect.(\ref{end}) we briefly conclude.

\subsection{The de Sitter inner core}
As already mentioned in the introduction, all known regular BH solutions  exhibit a de Sitter geometry near the origin.
Thus, it is important to see what happens  in the case of Rastall gravity in the same region.\\
The short distance behavior of the  $\gamma\left(\, 1/2\ ; r^2/4l_0^2\,\right)$ function is given by

\begin{equation}
 \gamma\left(\, 1/2\ ; r^2/4l_0^2\,\right)=  \frac{r}{l_0} +\frac{r^3}{6l_0^3}+\cdots \label{sdlim}
\end{equation}
 
and one easily we finds that the short distance form of the metric is

\begin{equation}
 -g_{00}=g_{rr}^{-1}= 1 -\frac{2MG_N}{\sqrt{\pi}l_0} - \frac{MG_N}{6\sqrt{\pi}l_0^3}r^2 +O(r^4)
\label{ds}
\end{equation}

which \emph{is}, again, ia a de-Sitter line element, characterized by an effective cosmological constant 
$\Lambda = MG_N/ 2\sqrt{\pi}l_0^3$.
The presence of the extra constant $-2G_N M/ \sqrt{\pi}l_0$ does \emph{not} alter the geometry and, simply, amounts to
a constant rescaling of the units of length and time. In other words,a general metric tensor of the form
$\eta_{\mu\nu}=Diag\left(\, -\lambda_1\ , \lambda_2\ ,\lambda_3\ ,\lambda_4\,\right)$ with constant components, 
\emph{is} the Minkowski metric and can be written with $\lambda_{(i)}=1$ by a simple rescaling of the units 
along the four axis. 
The same reasoning applies to the metric (\ref{ds}). On the other hand, it is well known that de Sitter space-time  
has a constant,  \emph{finite}, Ricci scalar  $R=-4\Lambda$\footnote{In this regard, it is useful to remark
a difference with respect to General Relativity. In Rastall gravity the vacuum field equations in the presence of a 
cosmological  constants read
\begin{equation}
 R_{\mu\nu}-\frac{1}{2}\left(\, R-2\Lambda \,\right) g_{\mu\nu} + \kappa\lambda g_{\mu\nu}R=0
\end{equation}

If $\kappa\lambda =1/2$ we obtain

\begin{equation}
 R_{\mu\nu}+\Lambda g_{\mu\nu}=0
\end{equation}
 
and $R=-4\Lambda$.  General Relativity one obtains  $R=+4\Lambda$. }. 

\subsection{Regular Ricci scalar  and the  tangential pressure} 
Let us give  a more in depth look at the energy momentum tensor components:

\begin{equation}
 T^\mu_\nu =Diag\left(\, -\rho\ , p_r\ , p_\perp\ , p_\perp\,\right) 
\end{equation}

 The state equation is

\begin{equation}
 \rho =-p_r
\end{equation}

and the tangential pressure $p_\perp$ is determined  in terms of the matter density $\rho$ as

\begin{equation}
 \frac{d p_\perp}{dr} +\frac{2}{r}p_\perp= \frac{2}{r}\rho \label{ptang2}
\end{equation}

This equation must be solved with a boundary condition in $r=0$ consistent with (\ref{ds}) de Sitter metric, i.e.
\begin{figure}[h!]
\begin{center}
\includegraphics[width=8cm]{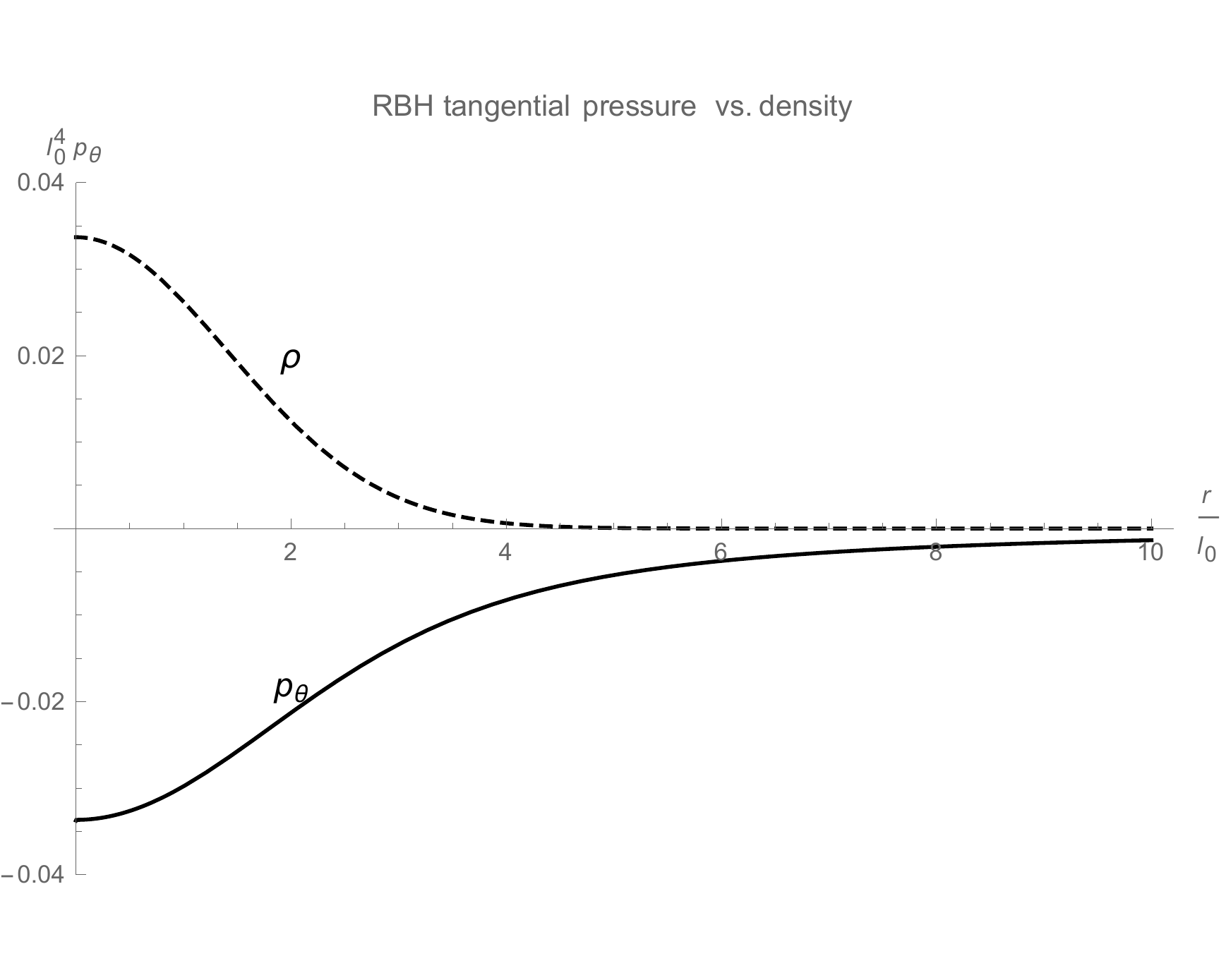}
\caption{Plots of the two functions (\ref{sorg}),(\ref{ptg}). Both of them are regular everywhere. }
\end{center}
\end{figure}
\begin{equation}
 p_\perp(0)=p_r(0)=-\rho(0)
\end{equation}

Thus, we the correct tangential pressure turns out to be

\begin{equation}
 p_\perp(r)=\frac{M}{2\pi^{3/2}l_0 r^2}\left[\, e^{-r^2/4l_0^2}-1\,\right]\label{ptg}
\end{equation}

Equation (\ref{ptg}) is finite everywhere, vanishes as $r\to\infty$, and satisfies short distance behavior

\begin{equation}
 p_\perp(0)=-\frac{M}{(4\pi)^{3/2}l_0^3}=-\rho(0)< \infty
\end{equation}

In fact, considering short distance behavior of the Ricci scalar one finds

\begin{eqnarray}
&& R =2k\left(\, p_\perp -\rho\,\right)\ ,\nonumber\\
&& R(0)= 2k\left(\, p_\perp(0) -\rho(0)\,\right)= -4k\rho(0)=-4\Lambda <\infty 
\end{eqnarray}
This result asserts that the regular solutions of Rastall gravity behaves in the same way as the corresponding ones 
in  GR, as far as the properties of the curvature are concerned. No singular behavior appears whatsoever.

\subsection{Black hole evaporation}

Firstly, let us note that the mass spectrum of the Rastall Gaussian BHs has a lower bound. In fact, the mass-horizon
relation is given by

\begin{equation}
 M=\frac{r_H}{2G_N}\frac{\Gamma(1/2)}{\gamma\left(\, 1/2\ ; r^2_H/4l_0^2\,\right)}\label{madm}
\end{equation}
\begin{figure}[h!]
\begin{center}
\includegraphics[width=10cm]{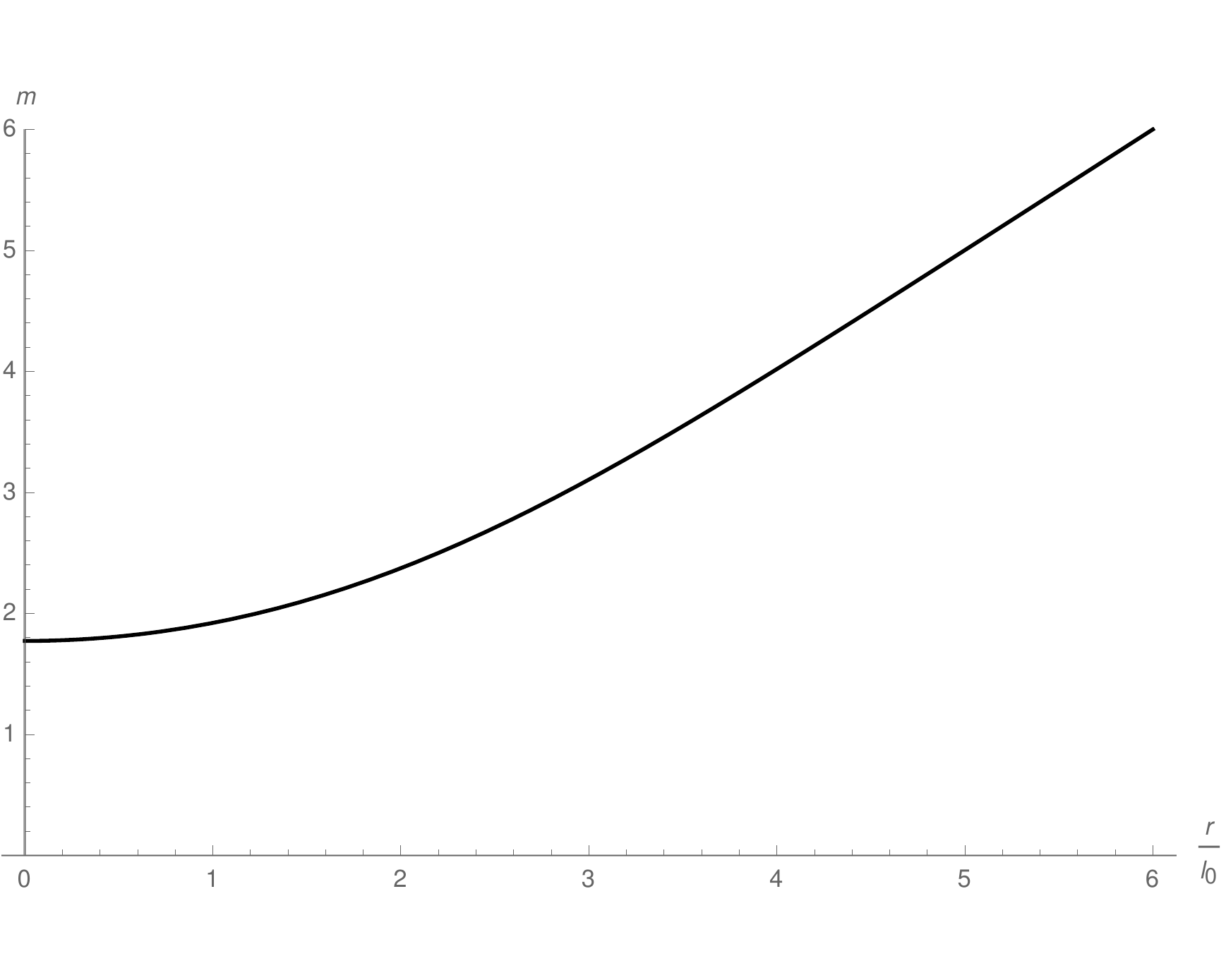}
\caption{Plot of the function (\ref{madm}). The intercept on the vertical axis represent the lower bound of the
BH mass spectrum.}
\end{center}
\end{figure}
This is a monotonically increasing function, for each $r_H\ge 0$ there is a unique value of $M\ge M_0$.  Below $M_0$ there
is no horizon and the line element (\ref{rastbh}) describes an \emph{ordinary} particle-like object. By using equation 
(\ref{sdlim}) one finds the value of minimal BH mass to be

\begin{equation}
 M_0=l_0 \frac{\sqrt\pi}{2G_N}=  \sqrt\pi M_{Pl}\frac{l_0}{l_{Pl}} \label{mmin}
\end{equation}
where, we use the standard definition of Planck length and mass, given by $2G_N=l_{Pl}^2= M_{Pl}^{-2}=l_{Pl}/M_{Pl}$ 
\footnote{The use of
the definitions $2G_N=l_{Pl}^2= M_{Pl}^{-2}=l_{Pl}/M_{Pl}$ guarantees that, in natural units $h=1\ ,c=1$, for $M=M_{Pl}$ 
the Schwarzschild radius and the Compton length coincide. }
The mass spectrum (\ref{madm}) is bounded from below by the minimum value $M_0$ needed to have an event horizon. 
Equation (\ref{mmin}) shows the relation between $M_0$ and the width $l_0$. Two choices of $l_0$ can be useful to
clarify the physical meaning of $M_0$. The most intuitive assignment is the identification $l_0 = l_{Pl}$, leading to

\begin{equation}
 M_0=\sqrt\pi M_{Pl} = 1.76\, M_{Pl}
\end{equation}
giving the same minimal mass as in Figure(1) .\\
However, one could argue that this choice is more appropriate in a full quantum gravity framework rather than in the
semi-classical regime we considering here. A more precautionary value for 
the width of the Gaussian distribution is
the Compton wavelength associated to the mass $M$ itself \cite{Spallucci:2017aod}. In this case

\begin{equation}
 M_0= \pi^{1/4}M_{Pl}=1.33 M_{Pl}\ ,
\end{equation}

which is even closer to the Planck mass. Thus, for any value of $l_0$ larger than the Planck length itself,
there is no way to produce a ``sub-Planckian'' black hole. May be that in some proper extension of Rastall gravity,
including large extra dimensions, the production threshold can be lowered enough to be in the range of next generation
of particle colliders \cite{Nicolini:2011nz,Mureika:2011hg,Nicolini:2013ega,Wondrak:2016urr,Wondrak:2017ttq}, 
but in the present minimal model the energy scale is still too high. \\
The Hawking temperature is given by
\begin{equation}
 T_H =\frac{1}{4\pi r_H}
\left[\, 1 -\frac{r_H}{l_0}\frac{e^{-r_H^2/4l_0^2}}{ \gamma\left(\, 1/2\ ; r^2_H/4l_0^2\,\right)  }\,\right]
\label{rht}
\end{equation}
\begin{figure}[h!]
\begin{center}
\includegraphics[width=8cm]{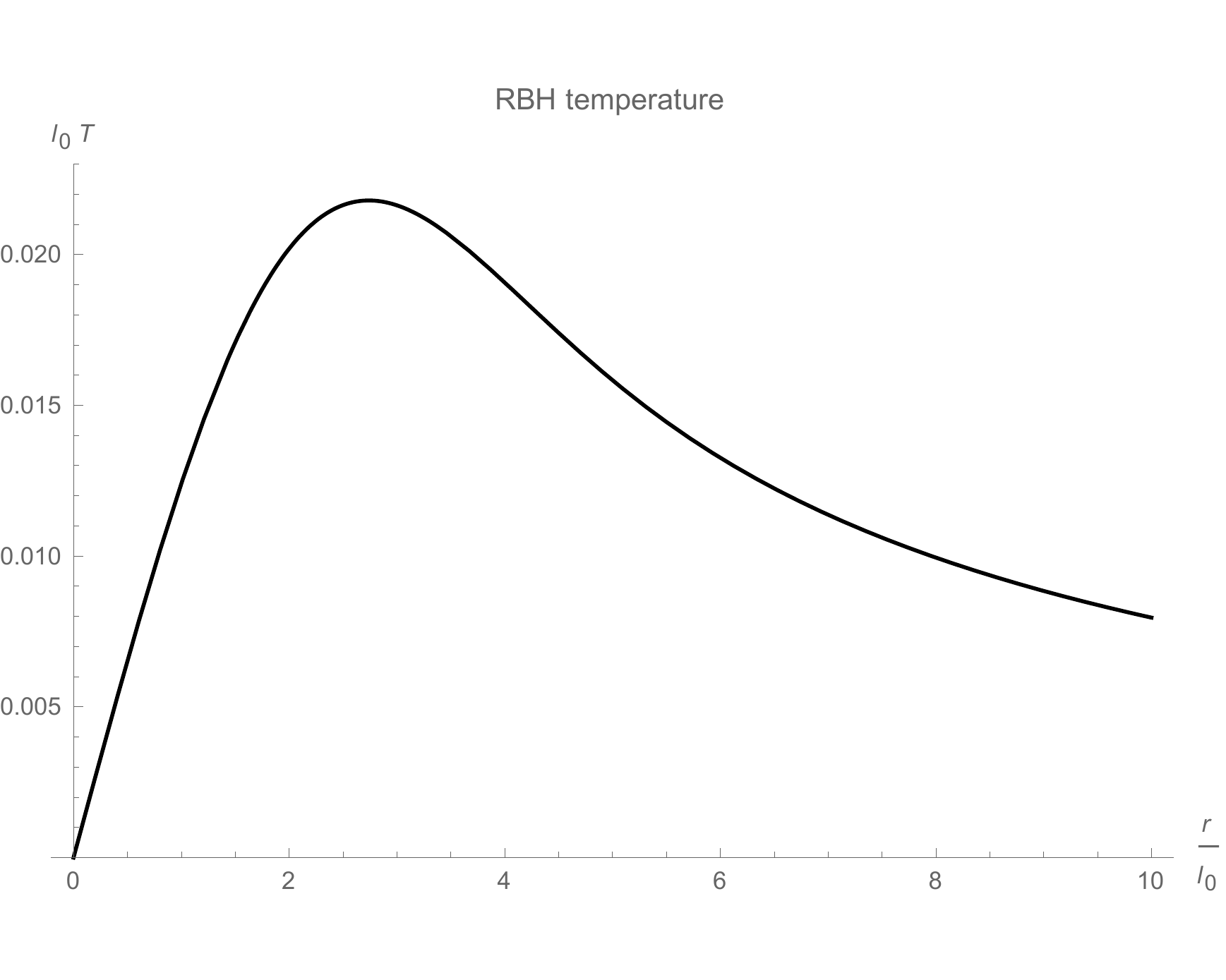}
\caption{Plot of the temperature (\ref{rht}). $T_H=0\leftrightarrow r_H=0$, i.e. the horizon completely evaporates
as $M\to M_0$. }
\end{center}
\end{figure}

It is interesting to describe the  final stage of evaporation of Rastall BH. As  $r_H\to 0$  temperature drops to zero

\begin{equation}
 T_H \simeq \frac{5r_H}{48\pi l_0^2}\to 0
\end{equation}

But, this occurs when $M\to M_0$. Thus, when $T_H$ reaches zero the horizon is completely evaporated away \emph{leaving} a 
frozen,particle-like remnant of mass $M=M_0$ dismissing claims in \cite{Ma:2017jko}. It is also interesting to notice 
that the total amount of thermal energy radiated via the Hawking process is

\begin{equation}
 E_{RAD}=M-M_0
\end{equation}

The frozen remnant is a regular, stable, massive lump of matter.\\
Another thermodynamical aspect of BH's is its entropy.
A quick integration of the ``first law'' gives the BH entropy as a function of the horizon radius
\begin{equation}
 S= \frac{\pi^{3/2}}{G_N}\frac{r_H^2}{\gamma\left(\, 1/2\ ; r^2_H/4l_0^2\,\right) }+ \frac{\pi^{3/2}}{G_Nl_0}
\int_0^{r_H}du \frac{u^2\,e^{-u^2/4l^2_0}}{\left[\,\gamma\left(\, 1/2\ ; u^2/4l_0^2\,\right)\,\right]^2}
\end{equation}

For \emph{large} BHs the entropy is given by the canonical area law $A_H/4$ \emph{plus} a small correction keeping 
the memory  of the extended nature of the source

\begin{equation}
 S 
\simeq \frac{\pi r_H^2}{G_N} +\frac{2\pi^{3/2} l_0^2}{G_N}
\end{equation}

In the opposite limit of \emph{small}, cold, BHs, we find
a linearly vanishing entropy, as it would be expected from the ``third law'', 

\begin{equation}
 S 
\simeq \frac{2\pi^{3/2} r_H l_0}{G_N}
\end{equation}

Again, regular BH solutions do not satisfy simple area law but receive corrections due to the extended character 
of the matter source. Rastall BH is no exception to this rule.

\section{Conclusions}
\label{end}
In this note we have shown that the Rastall gravity BHs behave in the same way as all known regular BH solutions, 
with the only difference that it exhibits a {\bf single} horizon. 
The Gaussian ``\emph{dirty}''-type BH, originally introduced in  \cite{Nicolini:2009gw},
which has been recently used in order to obtain regular metric, is not necessary in the Rastall framework.
We have shown  that a singularity free BH can  be obtained following standard procedure leading to other regular solutions 
in General Relativity, without the need  to set on a more ``~exotic~'' path as in \cite{Ma:2017jko}.\\
The choice $2\kappa\lambda = 1$ was dictated by the fact that (\ref{E1}) can be solved exactly. However, for any $\kappa\lambda\ne 1/4$
the conclusions of this paper still hold. As it has been shown in (\cite{Nicolini:2017moq}), whenever the energy momentum
tensor describes a regular matter/energy distribution the solution of the filed equations cannot lead to a singular metric. This property
is independent from the actual value of the numerical constant multiplying the Ricci scalar.

\end{document}